\documentclass{iaus}
\usepackage{graphicx}
\newcommand{\kms}{$\rm{\,km \,s}^{-1}$\ }
\newcommand{\MBH}{\ensuremath{M_\mathrm{BH}}}
\newcommand{\sigstar}{\ensuremath{\sigma_\star}}

\newcommand{\Msph}{\ensuremath{M_\mathrm{sph}}}
\newcommand\arcdeg{\mbox{$^\circ$}}% 
\newcommand{\Hubble}{\ensuremath{\mathrm{H}_0}}
\newcommand{\MPC}{\ensuremath{\mathrm{~Mpc}}}
\newcommand{\I}{\ensuremath{i}}
\newcommand{\1}{\ensuremath{^{-1}}}
\newcommand{\Lsph}{\ensuremath{L_\mathrm{sph}}}

\newcommand{\arcsec}{\ensuremath{\mathrm{arcsec}}}
\newcommand{\PC}{\ensuremath{\mathrm{~pc}}}

\title[Supermassive Black Holes in Spiral Galaxies]{Supermassive Black Holes in Spiral Galaxies}

\author[D. Macchetto]{Duccio Macchetto}
\affiliation{ESA and Space Telescope Science Institute,
       3700 San Martin Drive, Baltimore, MD 21218, USA;
	email: macchetto@stsci.edu}

\pubyear{2004}
\volume{222}
\pagerange{1--8}
\date{?? and in revised form ??}
\jname{The Interplay among Black Holes, Stars and ISM \\in Galactic Nuclei}
\editors{Th. Storchi Bergmann, L.C. Ho \& H.R. Schmitt, eds.}
\begin{document}

\maketitle
\begin{abstract}
We have embarked in an \emph{HST} program to determine the masses 
of black holes in spiral galaxies directly by measuring the line emission 
arising from an extended accretion disk. For each of the galaxies in our 
sample we have measured the rotation curve and determined the mass 
distribution within the inner 5--50~pc. We have modeled the stellar mass 
component using the photometric data from existing \emph{HST} images 
and using both data sets we have derived the masses of the black holes 
in each galaxy.   These results will be very important in clarifying the role 
of the black hole in powering the AGN, will shed light into the effectiveness 
of the accretion mechanisms and finally will be important in addressing the 
fundamental issue of unification for Seyfert~1 and Seyfert~2 galaxies.
\end{abstract}

%\keywords{black hole physics, galaxies: active, galaxies:
%bulges, galaxies: nuclei, galaxies: Seyfert}

\firstsection

\section{Introduction}
\label{intro}

It is widely accepted that Active Galactic Nuclei (AGN) are powered by
accretion of matter onto massive  black holes.  AGN activity peaked at
$z\sim1$--2   (e.g.,   Maloney   \&   Petrosian  1999)   and  the   high
($>10^{12}$L$_{\odot}$)  luminosities of  quasi stellar  objects (QSOs)
are  explained by  accretion onto  super massive  
($\sim10^8-10^{10}$~M$_{\odot}$)  black holes at or close to the Eddington 
limit.   The observed  evolution of  the space  density of  AGN (Chokshi  \& 
Turner 1992,  Faber et~al.\  1997, Marconi  \& Salvati  2001) implies that a
significant  fraction  of luminous  galaxies  must  host black  holes, relics of  
past activity.   Indeed,  it is  now  clear  that a  large fraction of hot spheroids 
contain a massive BH (e.g., Magorrian et~al.\ 1998;  van  der Marel  1999),  
and  it appears  that  the  BH mass  is proportional  to both  the  mass of the  
host  spheroid (Kormendy  \& Richstone  1995) and  its  velocity dispersion  
(Ferrarese \&  Merritt 2001, Gebhardt et~al.\ 2000, Merritt \& Ferrarese 2001).

Several radio-galaxies, all associated with giant elliptical galaxies, like M87 
(Macchetto et~al.\  1997),  M84 (Bower et~al.\ 1998), NGC~7052 (van der 
Marel \& van den  Bosch 1998) and Centaurus~A (Marconi et~al.\ 2001),  
are now known to host supermassive ($\sim10^8$--$10^9$~M$_{\odot}$)  
BHs in their nuclei. The luminosity of their optical nuclei indicates that they  
are accreting at a low rate and/or low accretion efficiency (Chiaberge, 
Capetti  \& Celotti  1999).   They presumably sustained  quasar activity in  
the past but at  the present epoch are emitting much below their Eddington  
limits (L/L$_\mathrm{Edd} \sim10^{-5}$--$10^{-7}$). The study of the 
Seyfert BH mass distribution  provides  a statistical method of investigating
the  interplay between  accretion  rate  and BH  growth.  In order  to
achieve this it is necessary to directly measure  the BH masses in Seyfert 
galaxies and to compare their Eddington and Bolometric luminosities using 
the hard X-ray luminosities. Similarly  important  will be  the comparison 
between the BH masses  found in Seyfert galaxies with those of  non active 
galaxies. However, to date, there are very few secure BH measurements 
or upper limits in spiral galaxies. It is therefore important to directly establish 
how common are BHs in spiral galaxies and whether they follow the same 
\MBH-\Msph, \MBH-\sigstar\ correlations as Elliptical galaxies.

To detect  and measure the masses of massive BHs requires spectral information 
at the highest possible  angular resolution---the ``sphere of  influence" of
massive BHs is typically $\sim 1''$ in radius even in the closest galaxies.  
Nuclear absorption line  spectra can be used to demonstrate the presence  
of a BH, but  the interpretation of the  data is complex because it involves 
stellar-dynamical models that have many degrees of freedom.  In  Seyfert 
galaxies  the  problems are compounded by the copious light from the AGN.   
Studies  at  \emph{HST} resolution  of  ordinary optical  emission  lines  from  
gas disks  in principle provide a more widely applicable and readily interpreted 
way of detecting BHs (cf.\ M87, Macchetto et~al.\  1997, Barth et~al.\  2001)
provided that the gas velocity field are not dominated by non gravitational motions.

\section{The Sample, Observational Strategy and Modeling} 
\label{sample}

Prompted by these considerations, we have undertaken a spectroscopic
survey of 54 spirals using STIS on the \emph{Hubble Space Telescope}.  Our
sample was extracted from a comprehensive ground-based study by Axon
et~al.\ who obtained H$\alpha$ and N\,\textsc{ii} rotation curves at a
seeing-limited resolution of $1''$, of 128 Sb, SBb, Sc, and SBc spiral 
galaxies from RC3. By restricting ourselves to galaxies with recession 
velocities $V<2000$\,km/s, we obtained a volume-limited sample of 
54~spirals that are known to have nuclear gas disks and span wide ranges 
in bulge mass and concentration.  The systemic velocity cut-off was 
chosen so that we can probe close to the nuclei of these galaxies, and 
detect even lower-mass black holes.  The frequency of AGN in our 
sample is typical of that found in other surveys of nearby spirals, with 
comparable numbers of weak nuclear radio sources and LINERS. 

The observational strategy, used for all the galaxies in our sample, consisted 
in obtaining spectra at three parallel positions with the central slit centered on 
the nucleus and the flanking ones at a distance of 0{\mbox{$.\!\!^{\prime\prime}$}}2.  
At each slit position we obtained two spectra with the G750M grating centered at 
H$\alpha$, with the second spectrum shifted along the slit by an integer number 
of detector pixels in order to remove cosmic-ray hits and hot pixels.  The nuclear
spectrum (NUC) was obtained with the 0{\mbox{$.\!\!^{\prime\prime}$}}1 slit and 
no binning of the detector pixels, yielding a spatial scale of 
0{\mbox{$.\!\!^{\prime\prime}$}}0507/pix along the slit, a dispersion per pixel of 
$\Delta\lambda = 0.554$~\AA\ and a spectral resolution of 
${\cal R} = \lambda/(2\Delta\lambda) \simeq6000$.  The off-nuclear spectra 
(POS1 and POS2) were obtained with the 0{\mbox{$.\!\!^{\prime\prime}$}}2 
slit and $2\times 2$ on-chip binning of the detector pixels, yielding 
0{\mbox{$.\!\!^{\prime\prime}$}}101/pix along the slit, 1.108~\AA/pix along 
the dispersion direction and ${\cal R}\simeq 3000$. The raw spectra were 
processed with the standard pipeline reduction software.  

We derived rotation curves for each of the observed slit positions and applied
our modeling code, described in detail in Marconi et~al.\ (2003), to fit the 
observed rotation curves.  Briefly the code computes the rotation curves 
of the gas assuming that the gas is rotating in circular orbits within a thin 
disk in the galaxy potential. The gravitational potential has two components: 
the stellar potential, determined from WFPC or NICMOS observations 
and characterized by its mass-to-light ratio, and a dark mass concentration 
(the black hole), spatially unresolved at \emph{HST}+STIS resolution and
characterized by its total mass M$_\mathrm{BH}$.  In computing the rotation 
curves we take into account the finite spatial resolution of \emph{HST}+STIS 
and we integrate over the slit and pixel area. The $\chi^2$ is minimized to 
determine the free parameters using a downhill simplex algorithm .
The emission line surface brightness is modeled with a composition
of two gaussians, the first reproducing the central emission peak
while the second accounts for the brightness behaviour at large radii.
Having fixed the line brightness distribution, the free parameters of the 
fit are the systemic velocity, $v_\mathrm{sys}$, the impact parameter (i.e., 
the distance between the slit center and the center of rotation) $b$, the 
position of the galaxy center along the slit $s_0$, the angle between the 
slit and the line of nodes, $\theta$, the disk inclination $i$, the mass to light 
ratio, $\Upsilon$ and the black hole mass $M_\mathrm{BH}$. We perform 
a $\chi^2$ minimization allowing all  parameters to vary freely. 

In this review I discuss the key results for two galaxies in our sample, 
namely NGC~4041 and NGC~5252, while in Fig.~1 I show a sample of the 
images and spectra for a subset of the other galaxies that we have observed. 

\begin{figure}%1
\bigskip
\centering
\includegraphics*[width=.6\linewidth]{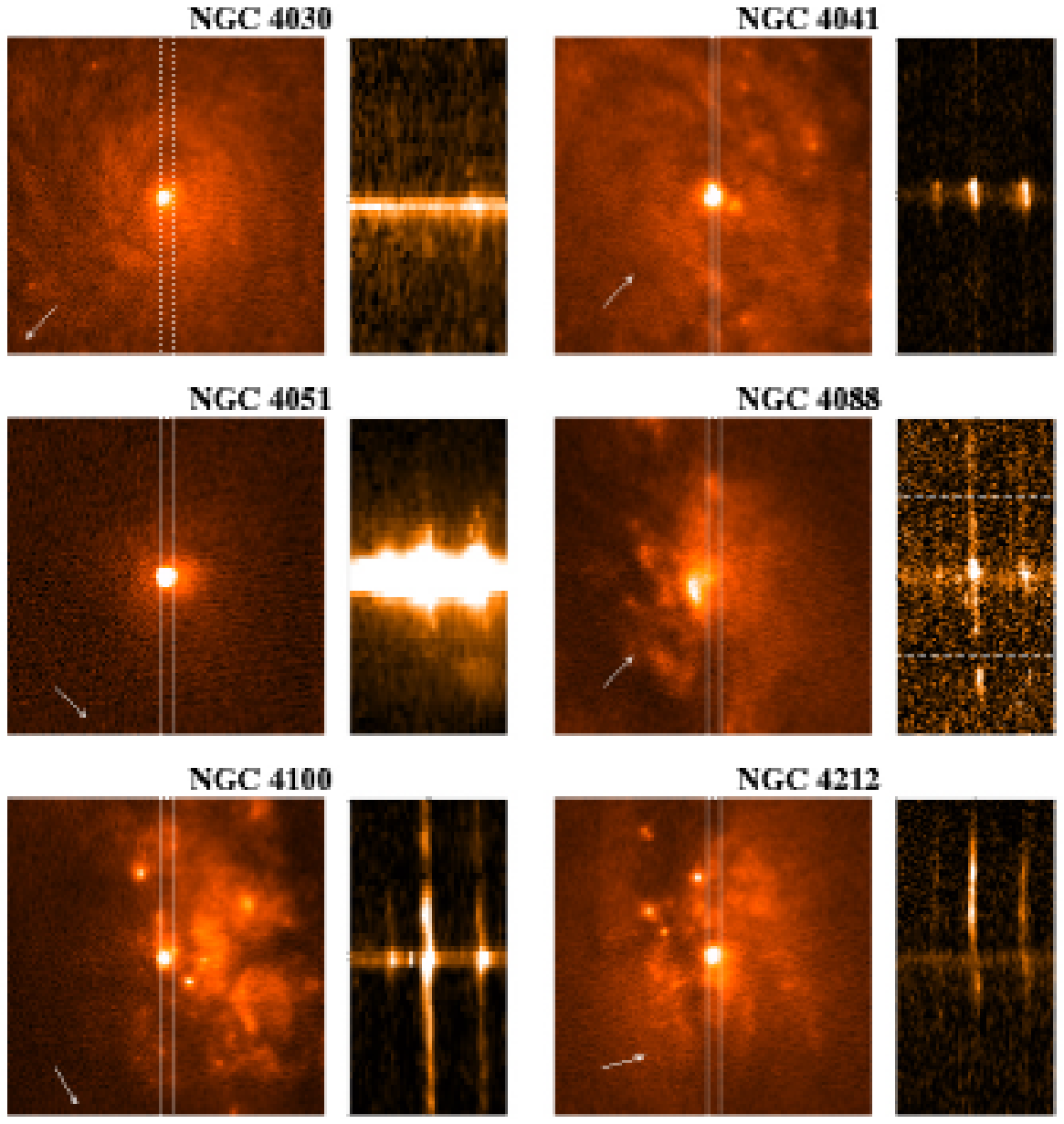}
\caption{}
\end{figure}

\section{NGC 4041} 
\label{NGC4041}

NGC 4041 is classified as a Sbc spiral galaxy with no detected AGN activity.
Its average heliocentric radial velocity from radio measurements is $1227\pm
9$\,\kms becoming $\simeq 1480$\,\kms after correction for Local Group
infall onto Virgo. With \Hubble=75\kms\MPC\1\ this corresponds to a
distance of $\simeq 19.5\MPC$ and to a scale of 95\PC/\arcsec.

\begin{figure}%2
\centering
\includegraphics[width=.4\linewidth]{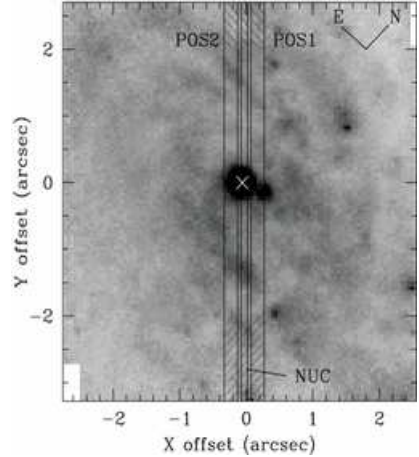}
\caption{\label{fig:acq} Slit positions overlaid on the acquisition
image. The 0,0 position is the position of the target derived from the
STIS ACQ procedure. The white cross is the kinematic center derived
from the fitting of the rotation curves.}
\end{figure}

The \emph{HST}/STIS spectra were used to map the velocity field of the 
gas in its nuclear region.  We detected the presence of a compact 
($r\simeq 0{\mbox{$.\!\!^{\prime\prime}$}}4\simeq 40\PC$), high surface
brightness, circularly rotating nuclear disk cospatial with a nuclear star
cluster. This disk is characterized by a rotation curve with a peak to peak
amplitude of $\sim 40$\,\kms and is systematically blueshifted by $\sim10$--20\,\kms 
with respect to the galaxy systemic velocity.

\begin{figure}%3
\centering
\includegraphics[width=.4\linewidth,angle=0]{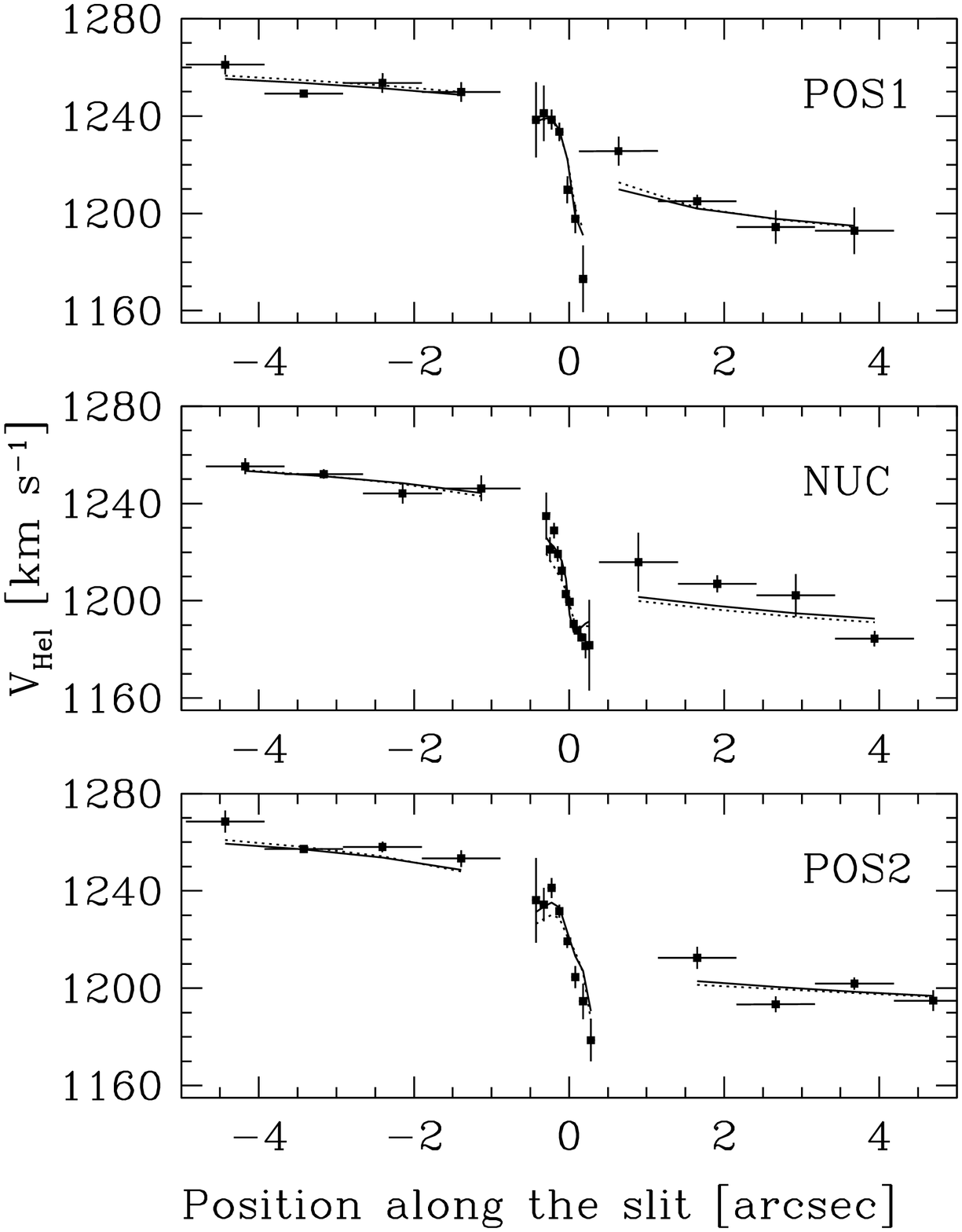}~\includegraphics[width=.4\linewidth,angle=0]{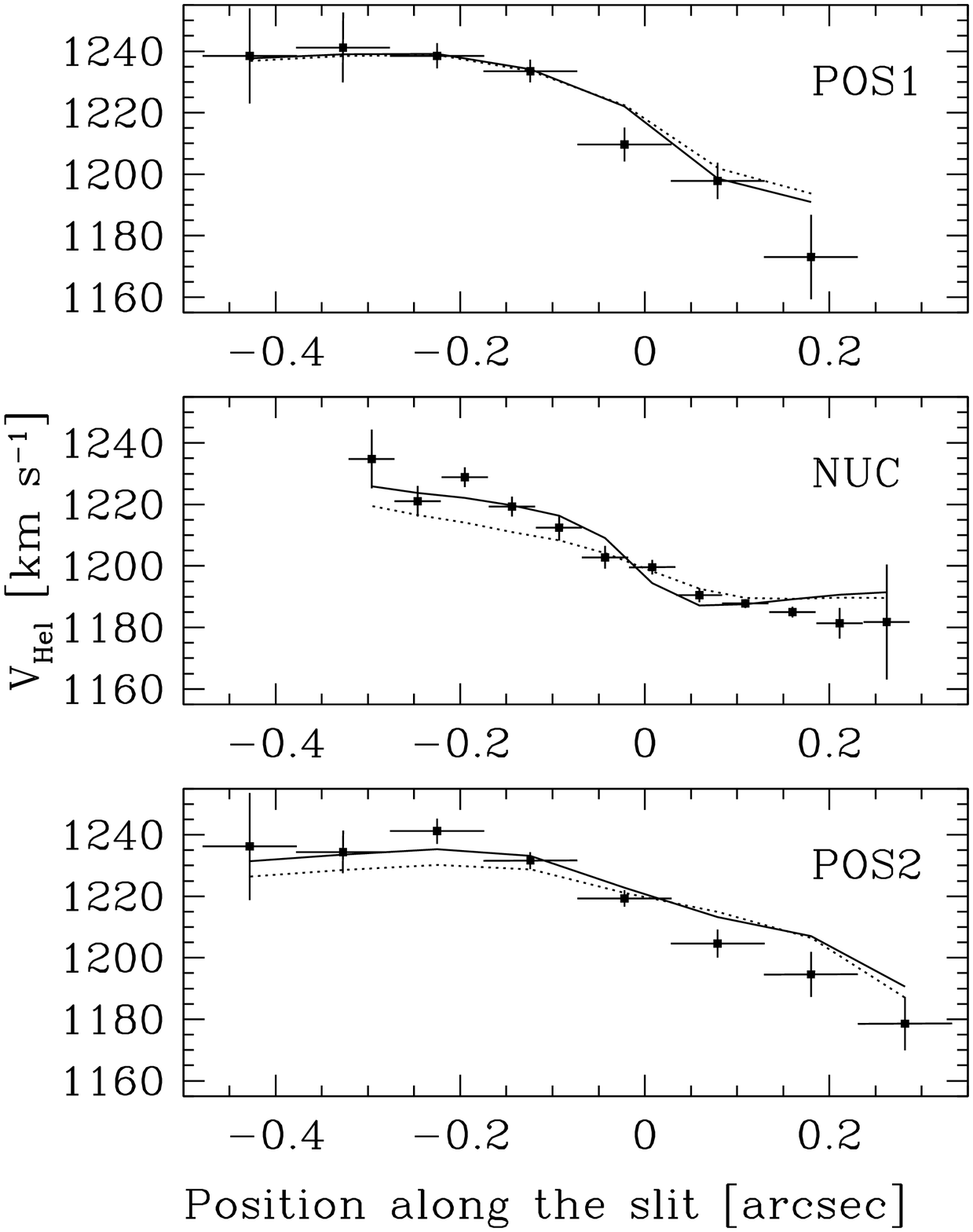}
\caption{\label{fig:fitstd} Best fit standard model of the observed rotation curves 
(solid line) compared with the data. The dotted line is the best fit model without 
a black hole. The model values are connected by straight lines in order to 
guide the eye. Note that points from external and nuclear regions are not
connected because they are kinematically decoupled.
The right panel is a zoom on the nuclear disk
region. }
\end{figure}

The standard approach followed in gas kinematical analysis is to assume
that (i)~gas disks around black holes are not warped i.e., they have the same
line of nodes and inclinations as the more extended components, and (ii)~the
stellar population has a constant mass-to-light ratio with radius
(e.g., van der Marel \& van den Bosch \1998; Barth et~al.\ 2001). Using the 
emission line flux distribution derived from the imaging data, the inclination of the 
galactic disk  can be fixed to \I =~20\arcdeg, i.e., The remaining free parameters 
can then be the fit with the procedure described earlier and we find that, in order to 
reproduce the observed rotation curve, a dark point mass [supermassive BH] of 
$(1_{-0.7}^{+0.6})\times10^{7}$~M$_{\odot}$ is needed. However, the blueshift of 
the inner disk suggests the possibility that the nuclear disk could be dynamically 
independent.  Following this line of reasoning we have relaxed the standard 
assumptions and model the curves by allowing variations in the stellar mass-to-light 
ratio and the disk inclination.  We have found that the kinematical data can be accounted 
for by the stellar mass provided that either the mass-to-light ratio is increased by a 
factor of $\sim 2$ or the inclination is allowed to vary.  This model resulted in a 
$3\sigma$ upper limit of $6\times10^{6}$~M$_{\odot}$ for the mass of any
nuclear black hole. 

Combining the results from the standard and alternative models, the present data 
only allow us to set an upper limit of $2\times10^{7}$~M$_{\odot}$ to the mass of the
nuclear BH.

If this upper limit is taken in conjunction with an estimated bulge B~magnitude 
of $-17.7$ and with a central stellar velocity dispersion of $\simeq
95$\,\kms, the putative black hole in NGC~4041 is not inconsistent with
both the \MBH-\Lsph\ and the \MBH-\sigstar\ correlations.

\section{NGC 5252} 
\label{NGC5252}

NGC 5252 is an early type (S0) Seyfert~2 galaxy at a redshift $z = 0.023$ whose 
line emission shows a  biconical morphology (Tadhunter \& Tsvetanov 1989) 
extending out to 20~kpc from the nucleus along  PA -15$^\circ$.  On  a sub-arcsec 
scale three emission line knots form a linear structure oriented at PA $\sim35$, 
close  to the bulge major  axis, suggestive of a  small scale gas disk. For  
H$_{\rm o} = 75$~Km~s$^{-1}$ Mpc$^{-1}$ at  the  distance  of 
NGC~5252  (92~Mpc),  $1''$ corresponds to 450~pc.

\begin{figure}%4
\centering
\includegraphics[width=0.4\linewidth]{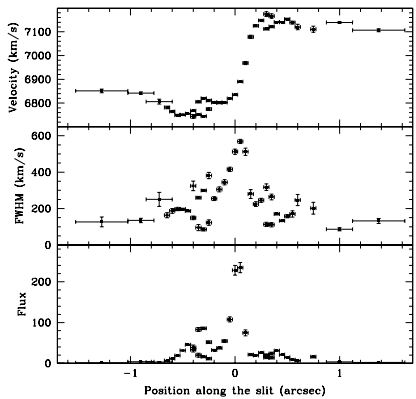}
\caption{\label{nuc} NGC5252.Velocity, flux and FWHM at the NUC (center) slit.}
\end{figure}

Fig.~\ref{nuc} shows  the line central velocity, flux and FWHM for the central slit 
of our STIS observations. Emission, which is detected out to a radius of 
$\sim1${\mbox{$.\!\!^{\prime\prime}$}}6 corresponding to $\sim720$~pc,  is 
strongly concentrated showing a bright compact knot cospatial with the continuum 
peak. Two secondary emission line maxima are also present at $\pm0.35''$ from 
the main peak. They represent the intersection of the slit with emission line knots 
seen in our  WFPC and STIS images. Two different gas systems are present in 
the nuclear regions of NGC~5252: the first shows a symmetric velocity field, with 
decreasing line width and can be interpreted as being produced by gas rotating 
around the nucleus. The second component, showing significant non circular 
motions, is found to be associated exclusively with the off-nuclear blobs.

Following the fitting procedure described earlier, the best fit to our data is obtained 
for a black hole mass  $M_\mathrm{BH} = 9.5\times10^8$~M$_{\odot}$. The model 
fitting of the nuclear rotation curve of NGC~5252  shows that the kinematics of gas 
in its innermost regions can be successfully accounted for by circular motions in a 
thin disk when a point-like dark mass (presumably a supermassive black hole) is 
added to the galaxy potential.

\begin{figure}%5
\centering
\includegraphics[width=0.35\linewidth]{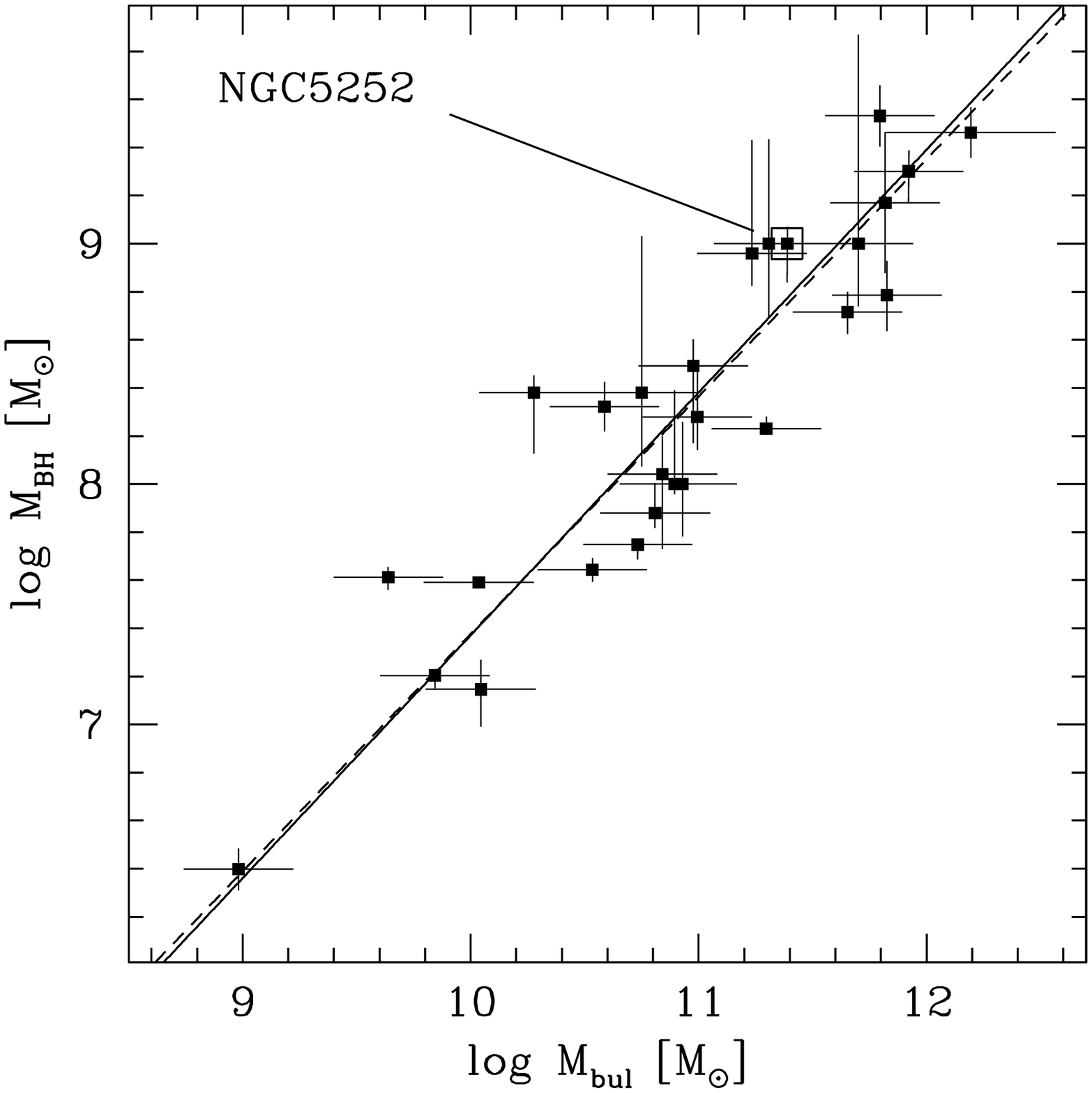}~\includegraphics[width=0.35\linewidth]{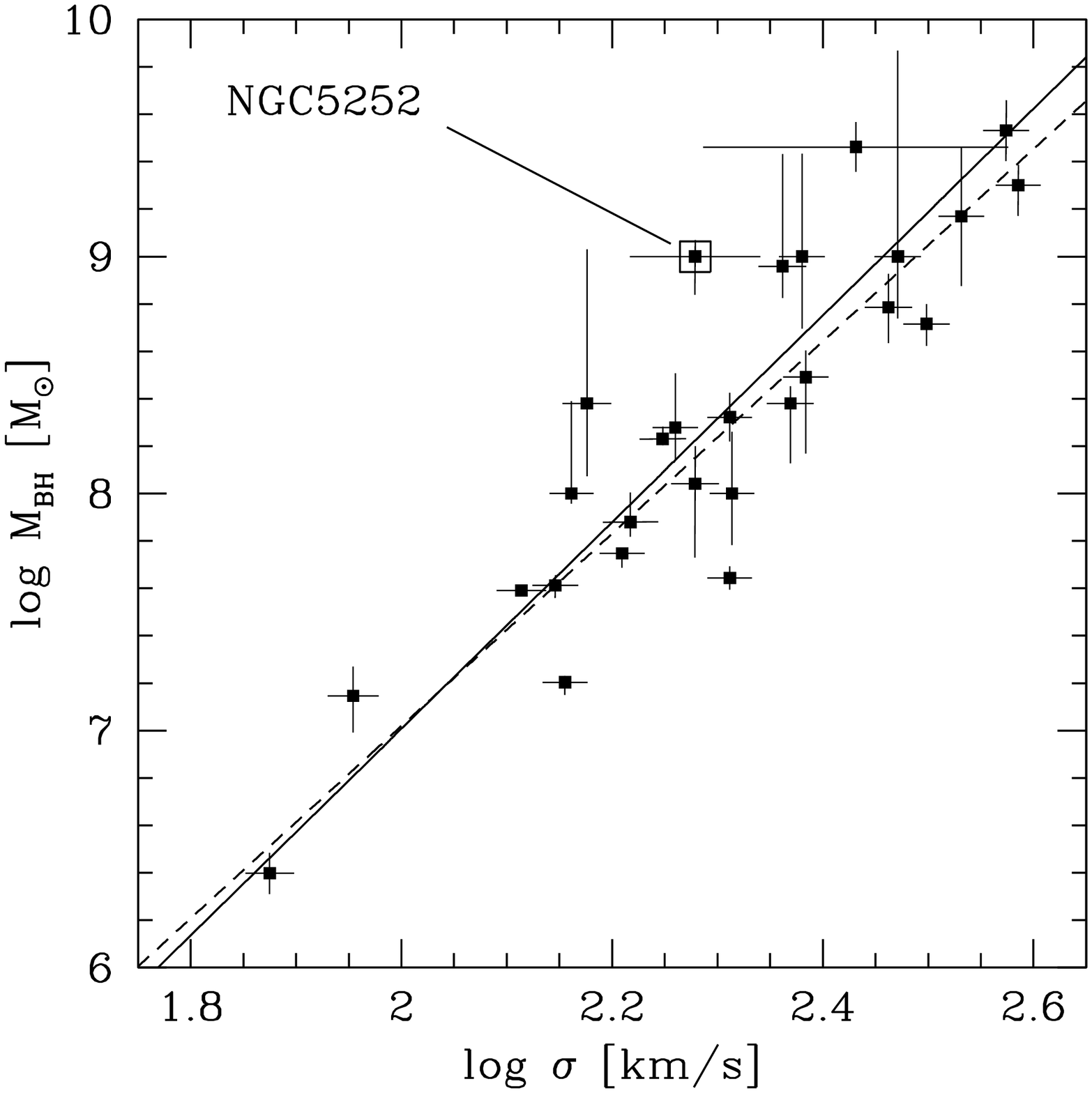}
\caption{\label{corr} Left, M$_\mathrm{BH}$ vs.\ bulge mass and right, M$_\mathrm{BH}$ vs.\ stellar velocity dispersion $\sigma_e$ with the best fits obtained from a bisector 
linear regression analysis (solid line) and ordinary least-square (dashed line).}
\end{figure}

The  central velocity  dispersion of  NGC~5252   (Nelson \& Whittle, 1995) is 
$190\pm27$~\kms. The correlation between velocity dispersion and black hole mass 
predicts a mass of $M = 1.0_{-0.5}^{+1.0}~10^8$~M$_{\odot}$ where the  error is 
dominated  by the uncertainty in  $\sigma_c$.  Therefore, the black hole mass we 
derived for NGC~5252  is larger by a factor $\sim10$ than the value expected  
from this correlation!  (see Fig.~\ref{corr}). This value however is in good 
agreement with the correlation between bulge and BH mass. As for its active nucleus, 
NGC~5252 is an outlier when compared to the available data for Seyfert galaxies, not 
only as it harbours a black hole larger than typical for these objects, but also as its host 
galaxy is substantially brighter than average for Seyfert galaxies. On the other hand,  
both the black hole and the bulge's mass are typical of the range for radio-quiet 
quasars. Combining the determined BH mass with the hard X-ray luminosity, we 
estimate that NGC~5252 is emitting at a fraction $\sim0.005$ of L$_\mathrm{Edd}$. 
This active nucleus thus appears to be a quasar relic, now probably accreting at a 
relatively low rate, rather than a low black hole mass counterpart of QSOs. 

\begin{acknowledgements}
The work described here is the result of a long standing collaboration with David Axon, 
Alessandro Capetti and Alessandro Marconi. They deserve the credit for the results.
\end{acknowledgements}

\end{document}